\documentclass[11pt,twoside,letterpaper]{article} %% The same for {book}
\usepackage{times,fancyhdr}
\usepackage[dvips]{graphicx}


\makeatletter 
\def\cleardoublepage{\clearpage\if@twoside \ifodd\c@page\else% 
    \hbox{}% 
    \thispagestyle{empty}%
    \newpage% 
    \if@twocolumn\hbox{}\newpage\fi\fi\fi} 
\makeatother

\def\figurename{Figure}
\makeatletter
\renewcommand{\fnum@figure}[1]{\figurename~\thefigure.}
\makeatother

\def\tablename{Table}
\makeatletter
\renewcommand{\fnum@table}[1]{\tablename~\thetable.}
\makeatother

\begin{document}
\title{
{\begin{flushleft}
%\vskip 0.45in
{\normalsize\bfseries\textit{Chapter~6}}
\end{flushleft}
%\vskip 0.45in
\vskip 0.1in
\bfseries\scshape A Closer Look at Gluons}}
\author{\bfseries\itshape Sadataka Furui\thanks{E-mail address: furui@umb.teikyo-u.ac.jp}\\
(formerly) Teikyo University, Graduate School of Science and Engineering\\
Utsunomiya, Tochigi, Japan}
\date{}
\maketitle
\thispagestyle{empty}
\setcounter{page}{1}
% ------- [First Page Running Head] - place it immediately after title! ------
\thispagestyle{fancy}
\fancyhead{}
\fancyhead[L]{In: Book Title \\ 
Editor: Editor Name, pp. {\thepage-\pageref{lastpage-01}}} % needs \label{lastpage-01} on the last page.
\fancyhead[R]{ISBN 0000000000  \\
\copyright~2007 Nova Science Publishers, Inc.}
\fancyfoot{}
\renewcommand{\headrulewidth}{0pt}
%------------------------------------------------------------------------------
Abstract: Gluons are strong interaction gauge fields which interact between quarks, i.e. constituents of baryons and mesons. Interaction of matters is phenomenologically described by gauge theory of strong, electromagnetic, weak and gravitational interactions.
In electro-weak theory, left handed leptons ${l}_L$ and neutrino ${\mathcal \nu}_L$, right handed leptons ${l}_R$ and left handed quarks  $u_L, d_L$ and right handed quarks $u_R, d_R$ follow $SU(2)\times U(1)$ symmetry. Charge of leptons and quarks define hypercharge $Y$, and via Higgs mechanism  $SU(2)_L\times U(1)_Y$ symmetry forms $U(1)_{em}$ symmetry. 

Presence of $J^P=3/2^+$ baryons, or $N^{*++}\sim uuu$ suggests a new degree of freedom ``color'' for quarks, which follows $SU(3)$ symmetry group. Hence the gluon fields are expressed as ${\mathcal A}^a_\mu$ where $a=1,2,\cdots, 8$ specify the color $SU(3)$ bases, and $\mu$ are 4-dimensional space-time coordinates. The quantum electrodynamics was extended to quantum chromo dynamics (QCD). Since $\partial^\mu {\mathcal A}^a_\mu$ is not a free field, the gauge theory requires ghosts that compensates unphysical degrees of freedom of gluons.
 Gluons, ghosts, leptons and quarks are related by Becchi-Rouet-Stora -Tyuitin (BRST) transformation of electro-weak and strong interaction of the $U(1)\times SU(2)\times SU(3)$ symmetric Faddeev-Popov Lagrangian. 

In order to describe Hadron dynamics properly, embedding of 4-dimensional space to 5-dimensional space was tried in lattice simulations, and in light front holographic QCD (LFHQCD) approach in which conformally symmetric light-front dynamics without ghost are embedded in $AdS_5$, and a parameter that fixes a mass scale was chosen from the Principle of Maximum Conformality. 
Coulomb or Landau gauge fixed Faddeev-Popov Yang-Mills field equation is known to have the Gribov ambiguity, and tunneling between vacua between different topological structures was proposed by van Baal and collaborators. 
The symmetry of three colors can be assigned three vectors of quaternion ${\bf H}$, whose multiplication on $2\times 2$ matrices of Dirac spinors on ${\bf S}^3$ induces transformations.  Instantons or sphalerons whose presence is expected from conformal equivalence of ${\bf S}^3\times {\bf R}$ to ${\bf R}^4$ are reviewed. An extension of the dynamics embedded in complex projective space is also discussed.

\vspace{0.1in}

\vspace{.08in} \noindent \textbf{Keywords:} Gluons, Ghosts, BRST Cohomology, Gribov-Zwanziger lattice simulation, LFHQCD, Sphalerons, Extra dimensions

%% Other situations:
%\noindent \textbf{Key Words}: Gluon, Color and Flavor degree of freedom 
\vspace{.08in} 
%\noindent {\textbf AMS Subject Classification:} %53D, 37C, 65P.

% ------------ [Running Heads - for odd and even pages] - please insert it only on page 2!
\pagestyle{fancy}
\fancyhead{}
\fancyhead[EC]{Furui, Sadataka}
\fancyhead[EL,OR]{\thepage}
\fancyhead[OC]{A Closer Look at Gluons}
\fancyfoot{}
\renewcommand\headrulewidth{0.5pt} 
%------------------------------------------------------------------------------

%++ Other useful packages to be used when needed ----------------------------------------------------
%\usepackage{epsfig}
%\usepackage{amssymb}
%\usepackage{amsmath}
%\usepackage{amsfonts}
%\usepackage{amsthm,amscd}
%\usepackage{amsbsy}
%\usepackage{latexsym}
%\usepackage{bm}
%\usepackage{url} 			%% Nicely format and linebreak URLs in the bibliography (and elsewhere).
%\usepackage{layout}
%\usepackage{pslatex}
%\usepackage{cite}
%\usepackage{fleqn} 		% displayed formulas flush left (default is centered).
%\usepackage{makeidx}
%\makeindex 						% Creates index at the end of the book (with makeidx).
%\usepackage{layout} 	% To see the current values of these dimensions, use the layout package, 
											% which defines a \layout command.
%\usepackage{epstopdf}	% Automatically converts EPS files to encapsulated PDF files (using ghostscript).
%\usepackage{color}
%\usepackage{hyperref}
%\usepackage[latin1]{inputenc}
%\usepackage[T1]{fontenc}
\raggedbottom                         %%% do NOT increase spaces between
                                      %%% paragraphs in order to fill always
                                      %%% the whole page
%\usepackage{poligraf} %% Color separation
%\usepackage[letter,cam,center]{crop} %% Printing crop-marks
%\usepackage{type1cm} 	%% Use scalable, PostScript Type 1 versions of the Computer Modern fonts.
%\usepackage{courier}	%% Replace the standard Computer Modern Typewriter font LaTeX uses
										 	%% for monospace text with the PostScript font Adobe Courier.
%\usepackage{lscape} 	% for landscape section
%\input tcilatex 			%% for Sci Word files
%++------------------------------------------------------------------------------

%\end{document}

\label{lastpage-01}
\newpage
\section{Introduction}

Feynman discusses on the problem of self-mass of charged particles in the section 28-4\cite{Feynman64} that “charges must be held to the sphere by some kind of rubber bands-something that keeps the charges from flying off.  It was first pointed out by Poincar\'e \cite{Poincare06} that the rubber bands -or whatever it is that holds the electron together-must be included in the energy and momentum calculation”. He presented that the difference of the electromagnetic mass $m^{(1)}=U_{elec}/c^2$ where $U_{elec}$ is calculated from $\epsilon_0 E^2$, where $E$ is the electric field produced by the static charge, and $m^{(2)}$ calculated from momentum derived from  the poynting vector $p=\frac{2}{3}\frac{e^2}{ac^2}\frac{v}{\sqrt{1-v^2/c^2}}$ are related by $m^{(2)}=\frac{4}{3}m^{(1)}$, and it was explained by Dirac\cite{Dirac38} as effects of self energy of the charged particle.
Feynman pointed out ambiguity in the self energy. 

In order to study interaction of hadrons, solving Dyson-Schwinger equations \cite{Dyson48, Dyson49, Schwinger49} and lattice simulations \cite{Creutz83} are two main methods. Dyson-Schwinger equation was used in analytical calculation of spectra of Quantum electro dynamics (QED), and applied to other fields. The lattice approach to quantum field theory  shows evidence that gauge theory is the fundamental tool, and that exchange of gauge gluons can confine quarks within subnuclear matter. 
In comparison to electrons or muons which have ${\bf Z}_2$ symmetry $(1,-1)$, presence of three quark system $N^{*++}\sim uuu$ suggested quarks have ${\bf Z}_3$ symmetry which is called color symmetry (red, green, blue), but experimentally the symmetry is not detected.

. 

In the minimal case of electromagnetism ${\bf Z}_2$ and gauge group $U(2)$\cite{Madore99}, fermions which take values in the space $M_2\otimes {\bf C}^2$, where $M_n$ is a $n\times n$ dimensional matrix, the {\it Maxwell} action is given by
\[
S_B=\frac{1}{4}\int F_{\alpha\beta}F^{\alpha\beta} dx,
\]
and the {\it Dirac} action is defined by a set of vectors $\tilde e_a$ on ${\bf S}^2$ that satisfy $\tilde e_a \tilde x^b-C^b_{ca}\tilde x^c$, where $C_{abc}=r^{-1}\epsilon_{abc}$.

Dual to the basis $\tilde e_a$ are components $\tilde\theta^a$ of the Maurer-Cartan form:
\[
d\tilde x^a=C^a_{bc}\tilde x^b \tilde \theta^c, \quad \tilde \theta^a=c^a_{bc}\tilde x^b d\tilde x^c-i A\tilde x^a.
\] 
Dirac proposed that the 1-form $A$ contains the potential of {\it Monopole}\cite{Dirac31,Ebert89} of unit magnetic charge and $F=dA$.
\[
S_F=Tr \int \bar\psi (\gamma^k D_k)\psi dx,
\]
where $\gamma^k=(1\otimes \gamma^\alpha,\sigma^a\otimes \gamma^5)$, 
\[
\gamma^0=\gamma_0=\left(\begin{array}{cc}
 I & 0\\
 0 & -I
 \end{array}\right), \quad
 -\gamma^k=\gamma_k=\left(\begin{array}{cc}
 0 & -\sigma_k\\
 \sigma_k & 0\end{array}\right),
\]
\[
\sigma_1=\left(\begin{array}{cc}
0 & 1\\
1 & 0\end{array}\right), \quad
\sigma_2=\left(\begin{array}{cc}
0 & -i\\
i & 0\end{array}\right), \quad
\sigma_3=\left(\begin{array}{cc}
1 & 0\\
0 & -1\end{array}\right),
\]
and \[
\gamma^5=\frac{1}{4!}\epsilon_{\mu\nu\alpha\beta}\gamma^\mu \gamma^\nu \gamma^\alpha \gamma^\beta.
\]

In the case of Yang-Mills field with connection $\omega$ and local section $\sigma$,  the gauge transformation can be expressed as
\[
\omega=h^{-1}A h +h^{-1} h
\] 
\[
A'=\sigma'^{*}\omega=g^{-1} A g +g^{-1} dg
\]
In terms of a moving frame on $V$
\[
F=\frac{1}{2}F_{\alpha\beta}\theta^\alpha\wedge \theta^\beta, \quad D\psi=D_\alpha\psi \theta^\alpha
\]

Schwinger \cite{Schwinger49, Schwinger83} showed that there could exist a conserved divergenceless electromagnetic stress tensor, which guarantees stability and covariance of the theory. He considered spherically symmetric charge distribution of total charge  $e$ at rest,
\[
\phi=ef(r^2),\quad {\bf A}=0,
\]
where $f(r^2)\sim (r^2)^{-1/2}$ and in uniformly moving rest frame 
\[
A^\mu(x)=\frac{e}{c} v^\mu f(\xi^2), \quad \xi^\mu=x^\mu+\frac{1}{c} v^\mu(\frac{1}{c} v x)
\]
where $v^2=v^\mu v_\mu=-c^2$ and $v\xi=0$.

Differential of vector fields are
\[
\partial^\nu A^\mu(x)=2\frac{e}{c}\xi^\nu v^\mu f'(\xi^2)
\] 
and
\begin{eqnarray}
\partial_\mu A^\mu(x)&=&2\frac{e}{c}\xi_\mu v^\mu f''(\xi^2)=0\nonumber\\
F^{\mu\nu}(x)&=&\partial^\mu A^\nu(x)-\partial^\nu A^\mu(x)=2\frac{e}{c}(\xi^\mu v^\nu-\xi^\nu v^\mu)f''(\xi^2)\nonumber
\end{eqnarray}

The current vector $j^\mu(x)$ produced by $\partial_\nu F^{\mu\nu}=\frac{4\pi}{c}j^\mu$ is
\[
j^\mu(x)=\frac{e}{2\pi}v^\mu[-2\xi^2 f'''(\xi^2)-3f'(\xi^2)].
\]
which is the covariant form of the rest frame charge density
\[
\rho=\frac{e}{2\pi}[-2r^2 f''(r^2)-3f'(r^2)]
\] 
whose total charge $\int_0^\infty dr r^2 4\pi\rho=e$ as calculated in\cite{Schwinger83}.

The rest mass based on the covariant form and also on action-principle approach to electromagnetic mass:  $m^{(1)}c^2$ and the electromagnetic mass  based based on the Poynting vector $\frac{1}{4\pi}({\bf E}\times {\bf B})$ defined as $m^{(2)}c^2$ are related by
\[
m^{(1)}c^2=\frac{3}{4} m^{(2)}c^2.
\]

Presence of Poincar\'e stresses to explain the electromagnetic mass of an electron $m_e$, the inertial mass $m_p$ and the effective mass $M=m_e+m_p$, that allow parametrization
\[
Mc^2=m_e c^2+m_p c^2=m_e c^2(1+(1+h)/3)=m_e c^2(4/3+h/3).
\]
where parameter $h=-1$ yields electrostatic mass $m^{(2)}$, and the parameter $h=0$ yields electrodynamic mass  $m^{(1)}$ was proposed by Jackson \cite{Jackson99}. The arbitraliness of the gauge fixing condition is discussed by  \cite{Rohrlich97, Rohrlich07, Medina06, Yaghjian06}.

The standard model with Higgs mechanism \cite{Higgs66} predicts that three of the original four $SU(2)\times U(1)$ gauge bosons become massive and one, corresponding to the photons, remains massless.  Although Schwinger did not suppose stress tensors of non electromagnetic origin, gluons could play a role \cite{Jackson99}, since three quarks can have electromagnetic charge and affect charged particles.

Experimentally, Dirac's magnetic monopoles \cite{Dirac48} are not observed, but 't Hooft \cite{tHooft74} and Polyakov \cite{Polyakov74, Polyakov87} showed in the Yang-Mills field theory a string-like topological singularity can appear.

In section 2, we explain the gluon degrees of freedom and its symmetries in the Yang-Mills theory. When one fixes the gauge symmetry of gluons in 4-dimensional Yang-Mills field theory, it is necessary to include ghosts to compensate unphysical degrees of reedom of gluons. Taking eigenstates of ghost number, one can formulate BRST symmetry.

The experimental detection of Higgs boson which appear by spontaneous symmetry breaking of the vacuum pushed investigation of the structure of the vacuum. In section 3, we explain the Gribov ambiguity that appears in the gluon propagator, and incorporation of instanton and sphalerons in QCD, using quaternion bases $\bf H$, and embedding ${\bf S}^3\times {\bf R}$ topology in ${\bf R}^4$ minkowski space.  

In section 4, we present a few formulations of hadron dynamics with additional dimension which corresponds to chiral symmetry or symmetry under Dirac's $\gamma_5$ operator.   The first formulation consists of light-front holographic QCD (LFHQCD) by Brodsky and collaborators\cite{dTDB15, Brodsky19} which is based on de Alfaro, Fubini and Furlan's anti de Sitter space approach \cite{dAFF76}.  The second formulation is based on non-commutative geometry developed by Connes and colaborators\cite{Connes94}. Non-commutative geometry is a set of tools to analyse broad range of objects that cannot be treated by classical methods \cite{WZ90}.  Non-commutative algebra appears when one incorporates Hamilton's quaternion $\bf H$ in the orthogonal bases of the Hilbert space\cite{Souriau70, Garling11}. Elements of $\bf H$ consist of ${\bf I}, {\bf i}, {\bf j}, {\bf k}$ which satisfy 
\begin{eqnarray}
{\bf i j}={\bf k},\quad {\bf j k}={\bf i},\quad {\bf k i}={\bf j}\nonumber\\
{\bf j i} =-{\bf k},\quad {\bf k j}=-{\bf i},\quad {\bf i k}=-{\bf j}\nonumber
\end{eqnarray}
In the $SL(2, {\bf C})$ representation, we can set
\[
{\bf I}=\left(\begin{array}{cc}
1  & 0 \\
0 & 1\end{array}\right),\quad
{\bf i}=\left(\begin{array}{cc}
0 & i  \\
i & 0 \end{array}\right),\quad
{\bf j}=\left(\begin{array}{cc}
0 & -1\\
1 & 0 \end{array}\right),\quad
{\bf k}=\left(\begin{array}{cc}
i & 0 \\
0 & i \end{array}\right).
\]
and 
\[
{\bf A}={\bf I}a+{\bf i}b+{\bf j}c+{\bf k}d=\left(\begin{array}{cc}
a+id  & -c+ib \\
c+ib & a-id \end{array}\right),\quad a,b,c,d\in{\bf R}
\]
and
\[
{\bf A}^*={\bf I}a-{\bf i}b-{\bf j}c-{\bf k}d.
\]
We define $a+i d=\alpha$ and $c+ib=\beta$, $\alpha,\beta\in {\bf C}$, and express
\[
{\bf A}=\left(\begin{array}{cc}
\alpha & -\bar \beta\\
\beta &\bar\alpha\end{array}\right), \quad
{\bf A}^*=\left(\begin{array}{cc}
\bar\alpha &  -\beta\\
\bar\beta &\alpha\end{array}\right).
\]

From algebra of complex numbers and quaternions, one can construct Clifford algebra \cite{Garling11} in quadratic spaces. In section 5, we explain Clifford Algebra on the manifold of ${\bf S}^3$ and embed ${\bf S}^3\times {\bf S}^1$ topology to complex projective space ${\bf CP}^3$.

Discussion on symmetry violation and topological approach in other physical phenomena is given in section6.

\section{Gluons and BRST Symmetry of Yang-Mills Field}
In Hamiltonian formulation of Yang-Mills field \cite{FS91}, lagrangian of QCD in the first order is written as
\begin{equation}
{\mathcal L}=-\frac{1}{2}tr[{\mathcal E}_k \partial_0 {\mathcal A}_k-\frac{1}{2}({\mathcal E}_k^2+{\mathcal G}_k^2)+{\mathcal A}_0{\mathcal C}],
\end{equation}
where 
\begin{eqnarray}
{\mathcal E}_k&=&{\mathcal F}_{k0}, {\mathcal G}_k=\frac{1}{2}\epsilon^{ijk}{\mathcal F}_{ji}, {\mathcal C}=\partial_k{\mathcal E}_k-g[{\mathcal A}_k,{\mathcal E}_k]\nonumber\\
{\mathcal F}_{ik}&=&\partial_k {\mathcal A}_i-\partial_i{\mathcal A}_k+g[{\mathcal A}_i,{\mathcal A}_k].
\end{eqnarray}
The generator ${\mathcal A}_k$ is decomposed in longitudinal and transverse comoponents
\[
{\mathcal A}_k={\mathcal A}_k^L+{\mathcal A}_k^T 
\]
where $\partial_k {\mathcal A}_k^T=0$ and 
\[
\partial{\mathcal A}_k^L=\partial_k{\mathcal B}(x), \quad {\mathcal B}(x)=\frac{1}{4\pi}\int\frac{1}{|x-y|}\partial_k{\mathcal A}_k(y) dy.  
\]
When the transverse component ${\mathcal E}_k^T(x)$ is defined as $\partial_k Q(x)$, the constraint on the longitudinal component is written as
\[
\Delta Q-g[{\mathcal A}_k,\partial_k Q]-g[{\mathcal A}_k, {\mathcal E}_k^T]=0.
\]
In the gauge ${\mathcal A}_0=0$, three dimensional expression of ${\mathcal F}_{ij}$ becomes
\[
{\mathcal F}_{ik}=\partial_k {\mathcal A}_i-\partial_i {\mathcal A}_k +g[{\mathcal A}_i,{\mathcal A}_k].
\]

 Madore\cite{Madore99} defined orthogonal moving frame $\theta^\alpha$, whose metric satisfy
\[
g(\theta^\alpha \otimes  \theta^\beta)=g^{\alpha\beta}   
\]
and structure equations of the torsion form $\Theta^\alpha$ and curvature form $\Omega^\alpha_\beta$ defined as
\begin{eqnarray}
\Theta^\alpha&=&d\theta^\alpha+\omega^\alpha_\beta\wedge \theta^\beta\nonumber\\
\Omega^\alpha_\beta=d\omega^\alpha_\beta+\omega^\alpha_\gamma\wedge\omega^\gamma_\beta
\end{eqnarray}
satisfy Bianchi identity:
\begin{eqnarray}
&&d\theta^\alpha+\omega^\alpha_\beta \wedge\Theta^\beta=\Omega^\alpha_\beta\wedge\theta^\beta\nonumber\\
&&d\Omega^\alpha_\beta+\omega^\alpha_\gamma\wedge\Omega^\gamma_\beta-\Omega^\alpha_\gamma\wedge\omega^\gamma_\beta=0
\end{eqnarray}

The Yang-Mills action in covariant gauge in ${\bf R}^n$ is defined as\cite{Madore99} 
\[
S[A]=\frac{1}{2}Tr\int F\wedge *F=\frac{1}{4}Tr\int F_{\alpha\beta} F^{\alpha\beta}dx
\]
where Hodge duality map $*$ of $\Omega^p(V)$ into $\Omega^{n-p}(V)$ is defined for $p\le n$ to the $(n-p)$ forms as
\[
*\theta_{\alpha_1\cdots \alpha_p}=\frac{1}{(n-p)!}\epsilon_{\alpha_1\cdots \alpha_p \alpha_{p+1}\cdots \alpha_n} \theta^{\alpha_{p+1}}\wedge\cdots \wedge \theta^{\alpha_n} .
\]
The vacuum Yang-Mills equations $D_\alpha F^{\alpha\beta}=0$ where
\[
F=\frac{1}{2}F_{\alpha\beta}\theta^\alpha \theta^\beta
\] 
satisfies the gauge transformation relations $F'=g^{-1}F g$. If the gauge group is abelian, the equatio is the Maxwell equation.

%Let $\lambda_a$ for $1\leq a\leq n^2-1$ be an anti-hermitian basis of $SU(n)$. The product $\lambda_a\lambda_b$ can be written in the form
%\[
%\lambda_a\lambda_b=\frac{1}{2}C^c_{ab}\lambda_c+\frac{1}{2}D^c_{ab}\lambda_c-\frac{1}{n}g_{ab},
%\] 
%where $g_{ab}$ are invariant inner products. 

%f there are no effects of Higgs field $\phi$, which manifests itself in the connection $\omega$ of curvature form $\Omega^1({\mathcal A})$ as
%\[
%\omega=A+\omega_v, \quad \omega_v=\theta+\phi, \quad A\in {\Omega^1}_h, \quad \omega_v\in {\Omega^1}_v.
%\] 
%Higgs potential $V(\phi)$ is given by $-\frac{1}{4}Tr(\Omega_{ab}\Omega^{ab})$ and 
%\[
%\Omega=\frac{1}{2}\Omega_{ab}\theta^a\theta^b, \quad \Omega_{ab}=[\phi_a, \phi_b]-C^c_{ab}\phi_c
%\]

Quarks and gluons have $SU(3)$ color degrees of freedom, but these degrees of freedom are not observed experimentally and models of color confinement were proposed.
Functionals of the QCD green function are given by Lagrangian $\mathcal L$ describing quarks and gluons, gauge-fixing Lagrangian ${\mathcal L}_{Fix}$ and Faddeev-Popov ghost Lagrangian ${\mathcal L}_{Ghost}$.

Faddeev-Popov ghosts are unphysical fields that compensate unphysical gluon degrees of freedom \cite{KO79, BBJ81, CL84, HT92, BR06}.

\begin{eqnarray}
{\mathcal L}&=&\frac{1}{4}{F^a}_{\mu\nu}(x) {F^{\mu\nu}}_a(x)+\bar\psi (x)(i\gamma^\mu D_\mu-M)\psi(x),\nonumber\\
{\mathcal L}_{Fix}&=&-\frac{1}{2}(C^a[{\mathcal A}_\mu])^2=-\frac{1}{2\xi}(\partial^\mu {A_\mu}^a(x))^2,\nonumber\\
{\mathcal L}_{Ghost}&=&\bar u^a(x)\partial^\mu(\partial_\mu \delta_{ab}-g f_{abc} {A_\mu}^c(x)) u^b(x).
\end{eqnarray}
The Becchi-Rouet-Stora-Tyutin (BRST) transformation in infinitesimal form is\cite{BRS75, BBJ81}
\begin{eqnarray}
\delta_s A^a_\mu (x)&=& -f^{abc} {A_\mu}^b (x)u^c(x) \delta\bar\lambda-\frac{1}{g}\partial_\mu u^a(x)\delta\bar\lambda\nonumber\\
\delta_s\psi (x)&=&-i T_a \psi(x) u^a(x)\delta\bar\lambda\nonumber\\
\delta_s\bar\psi (x)&=&i\bar\psi (x) T_a u^a (x)\delta\bar\lambda\nonumber\\
\delta_s u^a (x)&=&-\frac{1}{2} f^{abc} u_b(x) u_c (x)\delta\bar\lambda\nonumber\\
\delta_s \bar u^a (x)&=&\frac{1}{g {\sqrt\xi}}C^a[{\mathcal A}_\mu(x)]\delta\bar\lambda=\frac{1}{g\xi}\partial^\mu {A_\mu}^a(x) \delta\bar\lambda
\end{eqnarray}
where $\xi$ is the gauge parameter. 

 The covariant derivative of $\psi$ is with color indices $k,i$ and  $SU(3)$ color generator $t^C$,
\[
(D_\alpha \psi)_k=[\partial_\alpha {\delta_k}^l-i g {A_\alpha}^C {(t^C)_k}^l] (\psi)_k.
\]

Henneaux and Teitelboim \cite{HT92} defined the phase space $P$ and smooth phase space function $C^\infty(P)$. One restricts $P$ a surface $\Sigma$ and functions that vanish on $\Sigma$ form an ideal in $C^\infty(P)$ which is denoted as $\mathcal N$.

The local Lagrangian $L$ and action on variables $y^i$ are defined by
\[
S[y(t)]=\int_{t_1}^{t_2} L(y^i, \dot y^i,  \cdots y^{(k)_i}) dt
\]
where $y^{(k)_i}=d^k y_i/dt^k$,
Gauge transformations of arbitrary parameter $\epsilon$ are
\[
\delta_\eta y^i={S_A}^i \eta^A
\]
\[
\delta_\epsilon S=\frac{\delta S}{\delta y^i}\delta_\epsilon y^i =\frac{\delta S}{\delta y^i}R^i_\alpha \epsilon^\alpha
\]
and Noether's identity is
\[
\frac{\delta S}{\delta y^i}R^i_\alpha=0.
\]

Kugo and Ojima \cite{KO78,KO79} pointed out that the QCD Lagrangian is invariant under the BRS transformation, and the gauge fields ${A_\mu}^a(x)$, the auxiliary field $B^a(x)$, the covariant derivative of the ghost field $D_\mu c(x)$ , anti-ghost field $\bar c (x)$ necessarily have massless asymptotic fields which form the BRS-quartet. 
The physical space ${\mathcal V}_{phys}=|phys\rangle$ is specified as the one that satisfies the condition $Q_B|phys\rangle=0$, where 
\[
Q_B=\int d^3x [B^a D_0 c^a-\partial_0 B^a\cdot c^a+\frac{i}{2}g \partial_0 \bar c^a\cdot (c\times c)^a]
\]
and $(F\times G)^a=f_{abc} F^b G^c$.

The Noether current\cite{HT92} corresponding to the conservation of the color symmetry is
\[
g{J_\mu}^a=\partial^\nu {F_{\mu\nu}}^a+\{Q_B, D_\mu\bar c\},
\]
where the ambiguity by divergence of antisymmetric tensor shuld be understood, and this ambiguity is utilised so that massless contribution may be eliminated for the charge $Q_B$ to be well defined \cite{NF99, NF01}.
Denoting $g(A_\mu\times\bar c)\to {u^a}_b \partial_\mu\bar \gamma^b$, one obtains $A$ has a vanishing asymptotic value and
\[
\int e^{ip(x-y)}\langle 0|T D_\mu c^a(x)g(A_\nu \times \bar c)_b(y)|0\rangle dx=(g_{\mu\nu}-\frac{p_\mu p_\nu}{p^2})u^a_b (p^2)
\]
where $T$ means Dyson's time-ordering product operations that appear in perturbative Schroedinger functional approach\cite{Dyson48}.

Kugo and Ojima\cite{KO79} modified the Noether current for color charge $Q^a$ such that
\[
g{J'_\mu}^a=g J_\mu-\partial^\nu {F_{\mu\nu}}^a=\{Q_B, D_\mu\bar c\}^a
\]
and the condition for the massless component in the current $\{ Q_B, D_\mu\bar c\}$ is absent is $\delta_a^b+u_a^b=0$. 

In Landau gauge, Gribov region $\Omega$ is specified by the variation with respect to $g=e^\epsilon$
\begin{eqnarray}
\Delta||A^g||^2&=&-2\langle \partial A|\epsilon\rangle +\langle\epsilon \-\partial {\mathcal D}|\epsilon\rangle\nonumber\\
\Omega&=&\{ A|-\partial {\mathcal D}\geq 0, \partial A=0\}
\end{eqnarray}
We defined lattice link operator $U_{x\mu}=e^{A_{x,\mu}}$, where $A_{x,\mu}=-A_{x,\mu}^\dagger$ is the $SU(3)$ Lie algebra operator, and gauge transformation was chosen as
\[
e^{A^g_{x,\mu}}=g_x^\dagger e^{A_{x,\mu}} g_{x+\mu}
\]
Landau gauge was realized by choosing  $g=e^\epsilon$ and minimizing
\[
||A^g||^2=\sum_{x,\mu} Tr {A^g_{x,\mu}}^\dagger A_{x,\mu}^g
\]
The lattice covariant derivative $D_\mu(A)=\partial_\mu+Ad (A_\mu)$ in SU(3) simulation is given in \cite{NF99}.

Zwanziger\cite{Zwanziger90, Zwanziger91} developed a globally correct gauge fixing procedure. He showed in the  of minimal Landau or Coulomb gauge lattice simulation, gluon propagator is infrared vanishing 

The Faddeev-Popov determinant for gauge fixed fields $A$, which are hermitian Lie algebra gauge fields is
\[
FP(A)=-\partial_i D_i(A)=-\partial_i (\partial_i+i ad A_i)
\]
where $(ad X) Y=[X,Y]$. 

In Coulomb gauge ${\partial_i A}^i=0$, one considers $M\sim{{\bf S}^3}$ and
\[
I(g;A)=\int_M Tr(\{[g]A_i\}^2)=\int_M Tr(\{A_i + i g^{-1}\partial_i g\}^2).
\]  
For $h=e^X$ and using
\[
e^{-X}\partial_i e^X=\frac{1-exp(-ad X)}{ad X}(\partial _i X)=\partial_i X+\frac{1}{2}[\partial _i X,X]+\frac{1}{6}[[\partial_i X,X],X]+\cdots,
\]
one finds \cite{vanBaal92}
\[
I(e^X; A)=||A||^2-2i\int_M Tr(X\partial A_i)+\int_MTr(X^\dagger FP(A)X) +\frac{i}{3}\int_M Tr(X[[A_i,X],\partial_i X])+\cdots.
\]
The Gribov region $\Omega$ is defined as the set of transverse gauge fields for which $FP(A)$ is positive. By gauge transformation, a trajectory of minimum points of $FP(A)$ could bifurcate into two trajectories of minimum points \cite{vanBaal95}

Zwanziger \cite{Zwanziger94} defined fundamental modular region $\Lambda$ specified by
\[
\Lambda=\{A| ||A||^2=Min_g||A^g||^2\}, \quad \Lambda\subset \Omega
\]
and the ghost propagator which is an inverse of the Faddeev-Popov operator $\partial {\mathcal D}$\cite{FP67} as
\[
G(x-y)\delta^{ab}=\langle \frac{1}{-\partial {\mathcal D}_\mu(A)}\rangle.
\]
He defined lattice gauge covariant derivative $D_{\mu}^{ab}$ that operates on link variable $U_\mu(x)$ whose relation to $A_\mu^a(x)$ is
\[
A_\mu^a(x)t^a=\frac{1}{2}[U_\mu(x)-U_\mu^\dagger(x)]_{traceless}.
\]
In the `t Hooft double line representation \cite{Coleman85}, quarks are described by single line, gluons and ghosts are described by double line. The Faddeev-Popov tensor gives the ghost propagator 
\[
G_{\mu\nu}(x-y)\delta^{ab}=\langle D_\mu^{ac} D_\nu^{bd}\frac{1}{-\partial{\mathcal D}(A)^{cd}}\rangle
\]
and its Fourier transform is
\[
G_{\mu\nu}(\theta)=\sum_x G(x)_{\mu\nu}e^{-i\theta(x+\frac{1}{2}e_\mu-\frac{1}{2}e_\nu)}
\]
whose projection in the transverse direction is
\[
G_{\mu\nu}^{TT}=(g_{\mu\nu}-\frac{p_\mu p_\nu}{p^2})G_{\mu\nu}.
\]
The ghost number correspods to the number of branching of a gluon in double line expressions.

Baulieu and Zwanziger \cite{BZ01} replaced the gauge fixing in 4 dimensional space to 5-dimensional formulation and succeeded in restricting 4 dimensional gluon field in the Gribov region and decoupling ghost fields.
Zwanziger\cite{Zwanziger03} derived an equation similar to Kugo-Ojima confinement equation.  

Infrared vanishing of the gluon propagator was consistent with the renormalizable theory of 't Hooft \cite{tHooft06}.

 Although the BRST symmetry is violated on lattice simulations\cite{GS96}, samples that we took close to the fundamental moduler region showed $u\sim -0.7$ in quenched simulation \cite{NF01, FN04}, and $u\sim -1$ in unquenched Kogut-Susskind fermion configurations produced by the MILC collaboration \cite{FN06A,FN06B}.
 
 Verification of Kugo-Ojima coefficient using Dyson-Schwinger equation was investigated by Alkofer et al. \cite{AHS09, HAS10, WA18}. 

Modern formulation of the Faddeev-Popov S-matrix elements of Yang-Mills field theory in Landau gauge is
\[
\langle in|out\rangle\sim \int exp^{iS[B]}\Delta[B]\Pi_x \delta(\partial_\mu B^\mu(x)) dB(x)\int \Pi_x D\Omega(x)
\]
where the gauge group is expressed as
\[
B_\mu\to B_\mu^\Omega=\Omega B_\mu \Omega^{-1}+\epsilon^{-1}\partial_\mu\Omega \Omega^{-1}
\]
and the factor $\Pi_x  \delta(\partial_\mu B^\mu(x))$ symbolyzes that integral is performed over transverse fields, and $\Delta [B]$ is chosen such that
\[
\Delta[B] \int \Pi_x \delta(\partial_\mu B^\mu(x)^\Omega) d\Omega=const
\]
holds. The analysis was extended from the real Minkowski space to the complex Riemann space in \cite{AJPS97}.

\section{The Gribov Copy Problem and the Structure of Vacuum}
Serious ambiguities in Coulomb or Landau gauge fixed Faddeev-Popov Yang-Mills field equation were studied in connection with BRS transformation \cite{Gribov78, Fujikawa79}.  Gribov ambiguity and gauge fixing procedures are investigated in \cite{MN78}.  Spontaneous symmetry breaking of the gauge theory and prediction of a scalar bosons \cite{Higgs66, EB64, Kibble67}  motivated studies of topology of Higgs field\cite{AFG75}. 

Polyakov \cite{Polyakov87} proposed instanton in non abelian gauge theory, and
Dashen, Hasslacher and Neveu\cite{DHN74}  and Yaffe\cite{Yaffe89} proposed sphalerons, which in ${\bf S}^3\times {\bf R}$ described by
\begin{eqnarray}
F_{ij}&=&\partial_i A_j-\partial_j A_i+2\epsilon_{ijk} A_k+[A_i,A_j]\nonumber\\
&=&\frac{2i\epsilon_{ijk}\tau_k s^2}{(1+s^2+b^2+2s b\cdot n)^2},\nonumber
\end{eqnarray}
where the gauge vectors of instantons are
\[
A_0=\frac{is \vec b\cdot\vec\tau}{1+s^2+b^2+2s b\cdot n}, \quad A_i=-i\frac{s^2+s b\cdot n)\tau_j+s(\vec b\wedge\vec \tau)_j}{1+b^2+s^2+2s b\cdot n},
\]
where $\vec b=b\cdot \vec e$, $e^i_\mu$ is the dreibein and $s=\lambda e^t$.
The sphaleron potential is
\[
{\mathcal V}=-\frac{1}{2}\int_{S^3} Tr(F_{ij}^2)=\frac{48\pi^2(1+s^2+b^2)s^4}{(1+s^2+b^2)^2-4s^2 b^2)^{5/2}}.
\]
Take any path $\gamma$ connecting $A_i=0$ and $A_i=n\cdot\sigma \partial_i n\cdot\bar\sigma$ and determining maximum $\mathcal V_m(\gamma)$, one defines the sphaleron \cite{vBHD92, PvB94}.

A spharleron have a definite Chern-Simons number 
\[
Q(A)=\frac{1}{8\pi^2}\int_{S^3} Tr(A\wedge dA+\frac{2}{3}A\wedge A\wedge A)
\]
 in $SU(2)$-Higgs theory  \cite{Dunajski10}.  Higgs boson of a mass of 125 $GeV/c^2$ was detected at Large Hadron Collider at CERN, between 2011 and 2012.

For a $g\in \mathcal G$, where $\mathcal G$ is gauge configurations, $F_A(g)=I(g; A)$ is for generic $A$, Morse function \cite{Milnor63} on $\frac{g}{\mathcal G}$, where $g$ is the group of local gauge transformation \cite{vanBaal92}. Properties of $U(n)$ anti-self-dual connections $A$ on a $C^n$ bundle over a torus $T^4$ and relation to instantons i.e. self-dual solutions of the Yang-Mills equation\cite{AHDM78} were studied by Braam and van Baal \cite{BvB89}.  
van Baal and Hari Dass \cite{vBHD92} studied instanton and sphaleron in the $SU(2)$ Yang Mills gauge theory. 
They considered $x_\mu\in {\bf R}^4$, $r^2=x_\mu^2$ and $n_\mu=x_\mu/r$ and redefined time $t$ through
 $r=R  exp(t/R)$ such that
\[
dx_\mu^2=exp(2t/R)(dt^2+R^2 dn_\mu^2)
\]
where $dn_\mu^2$ represents the metric of a unit sphere of volume $2\pi^2$, and coordinates $x_\mu$ are represented by quaternions \cite{Dirac45, Lounesto01, Garling11, MGS14} $\sigma_mu=({\bf I}, i{\bf\tau})$ and anti-quaternion $\bar\sigma_mu=({\bf I}, -i{\bf\tau})$ as
\[
x=x_\mu\sigma^\mu=\bar x^\dagger, \quad \sigma_i=-\bar\sigma_i=i\tau_i, \quad \sigma_4=\bar\sigma_4=1. 
\]

Choosing $R=1$ and defining dreibein on ${\bf S}^3$, using `t Hooft symbols $\eta$ and $\bar \eta$\cite{tHooft76}
\begin{eqnarray}
\sigma_\mu \bar\sigma_\nu-\sigma_\nu\bar\sigma_\mu&=&2i\eta^a_{\mu\nu}\tau_a\nonumber\\
\bar\sigma_\mu \sigma_\nu-\bar\sigma_\nu\sigma_\mu&=&2i\bar\eta^a_{\mu\nu}\tau_a\nonumber
\end{eqnarray}
where $a=1,2,3$ and $\mu,\nu$ run from 1 to 4, as
\[
e^a_\mu=\eta^a_{\mu\nu} n_\nu.
\]

In the absence of fermions, the $\theta$ parameter is the relevant quantum number to connect the wave functionals in te various vacua. 
van Baal and Cutkosky \cite{vBC92} verified the tunnelling from the ${\bf A}=0$ vacuum to the vacua that have Chern-Simons number one or minus one in $SU(2)$ gauge.

${\bf S}^3\times {\bf R}$ is isomorphic to torus $T^4={\bf C}^4/G$ where $G$ is a discrete subgroup of a complex manifold $W$.

\section{Hadron Dynamics without Ghosts in Space-Time with an extra Dimension}
`t Hooft and Veltman \cite{tHV72A} pointed out that in 4-dimensional space, the operator
\[
\gamma^5=\frac{1}{4!}\epsilon_{\mu\nu\alpha\beta}\gamma^\mu \gamma^\nu \gamma^\alpha \gamma^\beta
\]
induces anomalies due to singularities of Feynman integral, but they can be cancelled to make the theory renormalizable \cite{tHV72B}.  The chiral flavor $SU(N)_L\otimes SU(N)_R\otimes U(1)$ currents
\begin{eqnarray}
J_\mu^{st}&=&i\bar\psi^s \gamma_\mu \psi^t\nonumber\\
J_\mu^{5 st}&=&i\bar\psi^s \gamma_\mu\gamma_5 \psi^t\nonumber
\end{eqnarray}
are conserved but
\[
J_\mu^5=\sum_t J_\mu^{5 tt} 
\]
has the Adler-Bell-Jackiw anomaly charcterized by
\begin{eqnarray}
\partial_\mu J_\mu^5&=&-i(Ng^2/16\pi^2)G_{\mu\nu}^a \tilde G_{\mu\nu}^a\nonumber\\
G_{\mu\nu}^a&=&\partial_\mu A_\nu^a-\partial_\nu A_\mu^a +g\epsilon_{abc} A_\mu^b A_\nu^c\nonumber\\
\tilde G_{\mu\nu}^a&=&\frac{1}{2}\epsilon_{\mu\nu\alpha\beta}G_{\alpha\beta}^a
\end{eqnarray}
and topological quantum number
\[
n=(g^2/32\pi^2)\int G_{\mu\nu}^a \tilde G_{\mu\nu}^a d^4x.
\]
`t Hooft linked isospin to one of $SO(3)$ subgroups of $SO(4)$\cite{tHooft76}.

Dosch, de T\'eramond and Brodsky \cite{DdTB15} showed that if one embeds 4-dimensional Euclidean space in 5-dimensional anti-de Sitter space\cite{PR84,PR86} by adopting the AdS/CFT corespondence \cite{Mardacena99}, and choosing the light-front holographic coordinate, one can construct QCD theory without ghost and get insight to the color confinement problem. 

The supersymmetry of the low energy hadron spectra was proposed in \cite{dAFF76, FR84}, and applied in the Principle of Maximum Conformality (PMC), which was successful in explaining supersymmetric hadron spectroscopy
\cite{Brodsky06, Brodsky18} and infrared suppression of gluon propagators.  In light-front quantum field theory, Pauli-Lubanski pseudovector 
\[
W^\mu=-\frac{1}{2} \epsilon^{\mu\nu\alpha\beta} P_\nu M_{\alpha\beta}
\]
where $P^2=m^2$ and $W^2$ are Casimir operators, and one can use light-front gauge in all Lorentz frames avoiding redundant gauge degrees of freedom characteristic of covariant gauges\cite{CB17}. 

Separation of quark spin degrees of freedom and gluon spin degrees of freedom of a nucleon and Lorentz symmetry 
in dynamics similar to light front dynamics is discussed in \cite{Ji98, Ji06}. 
He defined energy momentum of a nucleon $T^{\mu\nu}=T^{\mu\nu}_q+T^{\mu\nu}_g$ and total momentum operator $P^\mu=\int d^3 x T^{0\mu}$ 
The helicity operator $h=\vec J\cdot \hat P$ where $\hat p=\vec p/|\vec p|$ coincides with $W^\mu s_\mu$, where $s_\mu=(|\vec p|/M, p^0 \hat p/M)$. 

The choice of light-front time $\tau=t+z/c$ and light-cone coordinates $\xi_{\pm}=(\xi_0\pm \xi_3)/\sqrt 2$, and the transverse component $\xi_\perp=(\xi_1,\xi_2)$ allows Lorentz transformation proper conformal analysis.

\[
\langle p|P^\mu_{q,g}|p\rangle=A_{q,g}(\mu)p^\mu p+B_{q,g}(\mu)g^{\mu 0}/(2p^0).
\]
The angular momentum density $M^{\mu\alpha\beta}=T^{\mu\beta}x^\alpha-T^{\mu\alpha}x^\beta$ consists of quark and gluon parts
\[
M^{\mu\alpha\beta}_{q,g}=T^{\mu\beta}_{q,g}x^\alpha-T^{\mu\alpha}_{q,g}x^\beta.
\]
Gluon contribution to the angular momentum is the Poynting vector $J_g=\int d^3 x \vec E\times \vec B$.

In non-commutative geometry, one considers $n-$dimensional Euclidean vector space ${\bf C}^n$ with inner product, and interactions are descibed by $n\times n$ complex matrices $M_n$.
Let $V$ be a smooth compact oriented real manifold without boundary of dimension $m$, and ${\mathcal C}(V)$ be the commutative associative algebra of smooth real-valued function on $V$. Let $\partial_i$ be the natural basis of vectors on the space ${\bf R}^n$ and linear combination of $X=X^i\partial_i$ with  $X^i\in {\mathcal C}({\bf R}^n)$ modules and
\[
\chi ({\bf R}^n)=\oplus_1^n {\mathcal C}({\bf R}^n).
\]
Maxwell-Dirac action\cite{Madore99} can be defined by the algebra ${\mathcal A}={\mathcal C}(V)\otimes M_n$ where $M_n$ is the matrix algebra.

One lets ${\bf C}^n$ be the $n$-dimensional Euclidean vector space with the standard inner product, and for an integer $m<n$ write ${\bf C}^n={\bf C}^m\oplus {\bf C}^{n-m}$, and decompose $M_n$ as a direct sum 
$M_n={M_n}^+\oplus M_n^-$.  The $M_n^+$ are even operators that satisfy 
$M_n^+=M_m\times M_{n-m}\subset M_n$.

Yang-Mills action for $n=3$ and $m=1$ can be defined in 3 dimensional even matrices
\[
{\bf H}\oplus {\bf C}\subset {M_3}^+
\]
The total algebra is chosen to be ${\mathcal A}={\mathcal C}(V)\otimes {M_3}^+$ and modules of 1-form is
\[
\Omega^1({\mathcal A})={\mathcal C}(V)\otimes{\Omega^1}_\eta\oplus \Omega^1({\mathcal C}(V)) \otimes {M_3}^+
\]
where $\Omega^1({\mathcal C}(V))$ is the module of de Rham forms\cite{Madore99}.  
${\Omega^1}_\eta$ is ${M_3}^-$, i.e. polynomials of odd matrices of type
\[
\eta=\left(\begin{array}{ccc}
                         0 & 0 & a_1\\
                        0 & 0 & a_2\\
                        -a_1^*&-a_2^*&0\end{array}\right) .
\]
The model is called Connes-Lott \cite{CL90} model.

%General theory of $U(1)\times SU(2)\times SU(3)$ symmetry of non-commutative geometry was proposed by Connes\cite{Connes94}.
Application of noncommutative geometry to $SU(3)\times SU(2)\times U(1)$ symmetric standard model was done by Connes and Lott \cite{CL90}, and Chamseddine and Connes \cite{Connes94, CC97}. Noncommutative geometry appears when one incorporates quaternion in analysis, since for quaternions $p$ and $q$, results of multiplication $p\cdot q$ and $q\cdot p$ differes, if $p\ne q$. This additional degrees of freedom, or extra dimension was used to cover chiral-even and chiral-odd manifolds.

 The Lagrangian they took consists of
\begin{itemize} 
\item The pure gauge boson part ${\mathcal L}_G$,
\item The fermion kinetic term ${\mathcal L}_f$,
\item Kinetic terms for Higgs fields ${\mathcal L}_\phi$,
\item The Yukawa couplingof Higgs fields with fermion ${\mathcal L}_Y$,
\item The Higgs self-interaction ${\mathcal L}_V$.
\end{itemize}

For a compact Riemannian spin manifold $M$ and a self-adjoint Dirac operator $D=\partial_M$ acting in the Hilbert space $h$ of $L^2$ spinors on the manifold $M$, Connes et al constructed Yang-Mills action functional,
expressed as a triple $(\mathcal{A},  h, D)$, 
where $\mathcal A$ is the von Neumann algebra on $\mathcal M$:
\[
\mathcal A=C^\infty(M)\otimes {\mathcal A}_F
\]{
where ${\mathcal A}_F={\bf C}+{\bf H}+M_3({\bf C})$. Using a spectral geometry on ${\mathcal A}_F$ denoted as
$({\mathcal H}_F, D_F)$, the Hamiltonian and the Dirac operator are expressed as
\[
{\mathcal H}=L^2(M,S)\otimes {\mathcal H}_F, \quad D=i \gamma^\mu\partial_\mu M\otimes 1+\gamma_5\otimes D_F
\]
where $\partial^\mu _M$ means derivative on the manifold $M$.

They developed a theory of manifold in noncommutative geometry, using Hamiltonian algebra of quaternions $\bf H$ which is characterised by an element $x\in M_2({\bf C})$ that satisfies by $g\in U(2)$ and $J$ such that $JgJ^{-1}=\bar g$, the relation $J x J^{-1}=\bar x$.

In order to reproduce $U(1)\times SU(2)\times SU(3)$ algebra ${\mathcal A}={\bf C}\oplus {\bf H}$ and ${\mathcal B}={\bf C}+M_3({\bf C})$ were defined as $*$-algebra. 

The Hilbert space is
\[
h=h_0\oplus(h_1\otimes {\bf C}^3).
\]
The ${\mathcal A}$ bi-module structure is given for $\lambda\in{\bf C}$ and $q\in{\bf H}$
\begin{eqnarray}
(\lambda,q)(q_1,q_2)&=&(\lambda q_1, q q_2)\nonumber\\
(q_1,q_2)(\lambda, q)&=&(q_1 q, q_2\lambda)\nonumber
\end{eqnarray}
and 
\[
d(\lambda, q)=(q-\lambda,\lambda-q)\in {\bf H}\oplus{\bf H}, \quad (q_1,q_2)^*=(\bar q_1,\bar q_2)
\]
for $q\in {\bf H}$. A pair of differential forms $(F,G)\in \Omega_D^2$ consists of i) $(2,0)$ type: ${\bf C}$ valued 2-form $F$, and ${\bf H}$ vaued 2-form $G$, ii) $(1,1)$ type due to quaternonic 1-forms $(\omega_1, \omega_2)$,
iii) $(0,2)$ type due to a pair $(q_1,q_2)$. There is no ghosts.

The structure of the Clifford algebra of the finite space $F$ given by the linear map from the space ${\bf H}\oplus {\bf H}$ of 1-forms into the algebra of 1-forms $\sum_i a_i db_i$ and $\sum_i a_i[D, b_i]$, where 
\[
D=\left[\begin{array}{cc}
 0 & \partial\\
       \partial^*& 0
       \end{array}\right]
\]
where $\partial=\delta_1-i\delta_2$ and $\partial^*=-\delta_1-i\delta_2$ have different chilarity, remained as a problem\cite{Connes94}. The problem of $CP$ violation was discussed in \cite{CL90} following the model of Peccei and Quinn\cite{PQ77, Weinberg78}.

Orthogonal transformations in Euclidean $E^n$ space, $O(n)$ denotes the real orthogonal transformation, which is  the transitive group on the unit $(n-1)$ sphere ${\bf S}^{n-1}$.  If $x_0\in {\bf S}^{n-1}$ is fixed, subgroup leaving $x_0$ fixed is orthogonal group $O^{n-1}$, and we consider
\[
{\bf S}^{n-1}=O(n)/O({n-1}).
\]

One observes for $n=4$,
\[
SO(4)\simeq\frac{{\bf S}^3\times {\bf S}^3}{\{(1,1),(-1,-1)\}}
\]
or $M_4({\bf R})\simeq {\bf H}\otimes {\bf H}$ \cite{Lounesto01}.

In $SU(2)$ lattice QCD, a gauge transformation is written as 
\[
^U{A_\mu}=U A_\mu U^\dagger+U\partial_\mu U^\dagger,]
\]
where $U\in SU(2)$ is a mapping from ${\bf R}^3$ to ${\bf S}^3$.
%The $SU(2)$ group can be constructed on $U(1)$ bundle over ${\bf S}^2$  

%projected space $GL(n, {\bf C})/{\bf Z}_3$, where $GL(n,{\bf C})$ is the general linear transformation in $n$ complex number space and $/{\bf Z}_3$ means taking the quotient group \cite{Chevalley46}. One can replace the complex number $\bf C$ by quaternion $\bf H$ \cite{Chevalley46,Steenrod51}, whose base is composed of $e_0, e_1, e_2, e_3$, i.e. $q\in{\bf  H}$ is expressed in the form $\sum_{i=0}^3 a_i e_i$. 

In the case of $SU(3)$, using complex variables we consider $GL(2,{\bf C})$ transformation. 

\section{Clifford Algebra on the Manifold of ${\bf S}^1\times{\bf S}^3$} 
In Clifford algebra, finite dimensional division algebra ${\bf R}, {\bf C}$ and $\bf H$ can be treated equivalently.
Quaternions satisfy the division axiom - that the product of two factprs cannot vanish without either factor vanishing. We consider the Lorentz transformations of a quaternion $q\to q^*$ 
\[
q={\bf I}q_0+{\bf i}q_1+{\bf j}q_2+{\bf k}q_3
\]
that satisfy $q_0^2-q_1^2-q_2^2-q_3^2$ invariant. A transformation for complex numbers
\[
z^*=\frac{a z+b}{c z +d}
\]
which is general linear transformation is too general. We first choose $q=u v^{-1}$, $u,v\in \bf H$ \cite{Dirac45}.

With any quaternion $\lambda$
\[
q=u \lambda \lambda^{-1} v^{-1}=u\lambda (v\lambda)^{-1}
\]
To transform $q$ to $q^*$,  $u,v$ are transformed as
\[
u^*=a u+b v, \quad v^*=c u + d v
\]
where $a,b,c,d$ are arbitrary quaternions, and $q^*=u^* v^{* -1}$

By the above transformations
\begin{eqnarray}
q^*&=&(a u + b v)(c u +d v)^{-1}=(a u + b v)v^{-1} v (c u +d v)^{-1}\nonumber\\
&=&(a u + b v) v^{-1}[(c u+ d v)v^{-1}]^{-1}=(a q +b )(cq +d)^{-1}.
\end{eqnarray} 

Alternatively when $q=\alpha^{-1}\beta$
\[
q^*=(qa+b)^{-1}(q c+d).
\]
The transformation of previous assignment $q=u v^{-1}$ gives
\[
q^*=a(q+ a^{-1}b)[c (q+c^{-1} d)]^{-1}=a(q+a^{-1}b)(q+c^{-1} d)^{-1} c^{-1}.
\]
which yields a necessary condition $a^{-1}b\ne c^{-1}d$

The transformation $\tilde q=(a' q^*+b')(c' q^*+d')^{-1}=\tilde u \tilde v^{-1}$ and $\tilde q=q$  yields 
\begin{eqnarray}
&& a'b+b' d=0, \quad c'a +d' c=0\nonumber\\
&& a'a+b' c=c'b+d'd=m
\end{eqnarray}
where $m$ is a real number, not zero. The condition nan be satisfied by \cite{Dirac45}
\begin{eqnarray}
a'&=&\frac{m}{b^{-1}a+d^{-1}c} b^{-1}=\frac{m}{a-bd^{-1}c},\nonumber\\
b'&=&\frac{-m}{b^{-1}a-d^{-1}c}d^{-1}=\frac{m}{c-db^{-1}a},\nonumber\\
c'&=&\frac{m}{a^{-1}{b}-c^{-1}d}a^{-1}=\frac{m}{b-ac^{-1}d},\nonumber\\
d'&=&\frac{-m}{a^{-1}b-c^{-1}d}c^{-1}=\frac{m}{d-ca^{-1}b}.
\end{eqnarray}

In order to make $q, q^*$ and $\tilde q$ satisfy Lorentz group relations,
\[
Q_1=u\bar v, \quad Q_2=u\bar u, \quad Q_3=v\bar v
\]
are defined, and in $Q_1$, $u,v$ are replaced by $u\lambda, v\lambda$, which yields
\[
u\lambda (\bar{v\lambda})=u\lambda \bar\lambda\bar v=Q_1\lambda\bar\lambda
\] 
Put $Q_1={\bf I}X_0+{\bf i}X_1+{\bf j}X_2+{\bf k}X_3$, where $X_0, X_1,X_2, X_3\in {\bf R}$, and
\[
Q_2=X_4-X_5, \quad Q_3=X_4+X_5
\]
three $Q$s define six real numbers $X_0,X_1,X_2,X_3,X_4, X_5$. If $u,v$ are replaced by $u\lambda,v\lambda$, all $X$ are multiplied by $\lambda\bar\lambda$ and their ratios are not changed.

From relations $Q_1\bar Q_1=u\bar v v \bar u=u Q_3\bar u=Q_2 Q_3$. Hence
\[
X_0^2+X_1^2+X_2^2+X_3^3=X_4^2-X_5^2.
\]
By choosing $\tilde q$, one can define new $X$'s and to make the transformations satisfy the Lorentz group, one can impose restictions of choosing planes $X_0=0, \quad X_5=0$, which is equivalent to
\[
u\bar v+v\bar u=0, \quad u\bar u+v\bar v=0.
\]
The transformation that satisfy Lorentz group is
\[
q^*=(a q\pm \mu a)(-\mu \bar a q\pm a)^{-1}.
\]
Four quantities $\xi_\nu=X_\nu/X_5$ and $\eta_\nu=X_\nu/X_0$ $(\nu=1,2,3,4)$ transform as vectors in space-time, and we choose $\xi$ for analysis. A choice $u=q, v=1$ gives
\[
\xi_i=\frac{2 q_i}{1-|q|^2} (i=1,2,3), \quad \xi_4=\frac{1+|q|^2}{1-|q|^2}
\]
and
\[
{\xi_4}^2-{\xi_1}^2-{\xi_2}^2-{\xi_3}^2=1+\frac{4 q_0^2}{(1-|q|^2)^2}.
\]

The chiral symmetry in complex projective space ${\bf CP}^{N-1}$ was studied also by d'Adda et al. \cite{DALDV78, DADVL79}. They regarded quarks as Goldstone bosons and considered spontaneous symmetry breaking of $SU(N)$ to $SU(N-1)\otimes U(1)$ due to instantons.  Laglangian of $N^2-1$ scalar fields is defined as
\[
\mathcal L=\frac{1}{2}Tr\partial_\mu \phi \partial^\mu \phi-\lambda Tr P(\phi)
\]
where $P(\phi)$ is some polynomials\cite{Coleman85}. Choosing $N$ dimensional column vector of unit length $z^\dagger z=1$, $\phi=g_0^{-1}[N^{1/2}z z^\dagger-N^{-1/2}I]$, where $g_0$ is a parameter derived from $P(\phi)$.

After proper normalization,  and taking currents in the system as
\[
j_\mu=\frac{1}{2i}[z^\dagger\partial_\mu z-(\partial_\mu z^\dagger)z]
\]
the Lagrangian becomes
\[
{\mathcal L}=\partial_\mu z^\dagger \partial^\mu z-g_0^2 N^{-1}j_\mu j^\mu
\]
with the constraint $z^\dagger z=N/g_0^2$.

When $N=4$, the dimension of the real Goldstone fields becomes
\[
dim SU(4)-dim U(4-1)=15-(4-1)^2=6 
\]
which agrees with the dimension of $X$ fields.

%Only for $n=3$ and $n=7$, a bundle over ${\bf S}^n$ with group and fiber rotation group ${\bf R}^n$ which is subgroup of $O(n)$ is known to be equivalent to the product bundle ${\bf S}^n\times {\bf R}^n$
%We embed ${\bf S}^3\times {\bf S}^1$ in ${\bf R}^4$, where ${\bf S}^1$ can be used to define the background curvature as in Kaluza-Klein theory.  

Kodaira showed in projective space ${\bf P}^3$, the cohomology groups of ${\bf P}^3$ with coefficients in $\Omega({\bf C})$: $H^1({\bf P}^3,{\bf C})=0$\cite{Kodaira65} and one can define 2 algebraic curves on two dimensional surface. In projective space, zero-mode Dirac wave function contains a parameter that fixes the scale of the system. 

Kodaira defined a compex manifold $W$ as a domain ${\bf C}^2-(0,0)$, and a surface $S$ homeomorphic to ${\bf S}^1\times {\bf S}^3$ and $Z_t$ denotes an infinite cyclic group generated by a linea transformation $(z_1,z_2)\to (t z_1, t z_2)$ where $0< |t|<1$ and $S_t=W/Z_t$, and ${\bf S}^3$ is defined by $|z_1|^2+|z_2|^2=1$. He prooved existence of local complex analytical coordinates $(w_j, z_j)$ which define
\[
\theta=\partial/\partial z_j, \quad e^{c z_j}d w_j=\kappa^{m_j k}e^{c z_k} dw_k
\]
on $U_j\cap U_k$, where $\kappa$ is $0$ at $w_k=0$, and not zero elsewhere, $c$ is a constant, and $w_j$ is a holomorphic function of $w_k$ independent of $z_j$,
\[
w_j=f_{jk}(w_k)
\] 
on $U_j\cap U_k$.

Embedding ${\bf S}^1\times {\bf S}^3$ on which quarterion $\bf H$ operates on ${\bf S}^3$ in ${\bf R}^4$ and a parameter for the Laplace transform operates on ${\bf S}^1$ can be compared with conventional  embedding ${\bf S}^3\times {\bf R}$ in ${\bf R}^4$, since local complex  analytic coordinates $(w_j, z_j)$ defined respectively on $U_j$ such that the sheaf over $S$ is denoted as $\Theta$

%and the different structure of $CP$ symmetry violation and the sphaleron structure can be investigated.
%Color degrees of freedom of quark is  ${\bf Z}_3$ (Red, Blue and Green), and we adopt the Gupta-Bleuler scheme which drops unphysical degrees of freedom in physical expectation values, and consider the quatient space $SU(3)/{\bf Z}_3$ which  acts non-trivially on our system, which can be embedded to $SO(8)$ symmetry of orthogonal transformations of real numbers.
 % When one assigns the three degrees of freedom by $x_1\vec e_1, x_2\vec e_2, x_3\vec e_3$ on a three  dimensional sphere $S^3$ in 4 dimensional Euclidean space $R^4$ $(x_0,x_1,x_2,x_3)$, the point can be expressed by quaternion ${\bf H}=x_0+x_1\vec e_1+x_2 \vec e_2+x_3\vec e_3$, (${x_0}^2+{x_1}^2+{x_2}^2+{x_3}^2=1$). The unitary transformation of one quaternion $U(1,{\bf H})$ is isomorphic to $GL(2, {\bf C})$, where $\bf C$ means complex numbers. Qualitative differences of choosing smooth fuctions of QCD-Lagrangian on $S^1\times S^2$ and on $S^1\times S^3$ are studied.

Unitary transformation $U^n$ operating in complex $n-$space is also transitive on the unit $(2n-1)$ sphere, and we consider
\[
{\bf S}^{2n-1}=U(n)/U({n-1})
\]
and $n=2$. Using quaternion variables, we can consider $U(1,{\bf H})$ as the quotient group.

The group $SU(2)$ is a $U(1)$-bundle over ${\bf S}^2$ and the projection of the bundle onto its base space is called the {\it Hopf} fibration\cite{Madore99}. We extend the theory to bundles over ${\bf S}^3=O(4)/O(3)$. 
Hopf manifold is not algebraic manifold, but the Lie group $GL(2, {\bf C})$ can operate on $w_j$. One can choose two orbits corresponding to plus or minus $c$.

On Hopf surface $W={\bf C}^2-\{0\}$, one defines automorphisms 
\[
g_t: (z_1,.z_2)\to (\alpha z_1+t z_2, \alpha z_2), \quad 0< |\alpha |<1, t\in {\bf C}.
\]
$G_t=\{g_t^m|m\in {\bf Z}\}$ is automorphic and do not have fixed points, since
\[
g_t:^m (z_1,.z_2)\to (\alpha^m z_1+m \alpha^{m-1}t z_2, \alpha^m z_2).
\] 
$M_t={\bf C}/G_t$ is also a Hopf surface.
Complex structure of $M_0$ and $M_t, t$ not zero, are different. One dimensional algebraic manifold is called an algebraic curve. On $M_t, t\ne 0$, the regular linearly independent vector fields are $c_1(z_1\frac{\partial}{\partial z_1}+z_2\frac{\partial}{\partial z_2})$ and $c_2z_2\frac{\partial}{\partial z_1}$\cite{Kodaira92}. One can choose 2 algebraic curves on the Hopf surface. 
%For the canonical Hamiltonian $H_0(q,p)$ and the first class constraint $G_a\sim 0$ $(a=1,\cdots,\bar m)$, we define the extended Hamiltonian $H_E=H_0+\lambda^a G_a$ and 
%\[
%S_E[q^i(t), p_i(t), \Lambda^a(t)]=\int(\dot q^i p_i-H_0-\lambda^a G_a) dt.
%\]

%\[
%H_0(\delta)=\frac{(Ker \delta)_0}{(Im \delta)_0} =\frac{C^\infty(\Sigma)}{\mathca N}
%\]

%Ghosts $\eta^a$ are conjugate to the ghost momenta ${\mathcal P}_a$ 
%\[
%[{\matjhcal P}_a,\eta^b]=-(-1)^{(\epsilon_a+1)(\epsilon_b+1)}[\eta^b,{\mathcal P}_b]=-{\delta_a}^b
%\]

\section{Discussion and Conclusion}
We reviewed properties of Yang-Mills action, which is described by functions $F'=g^{-1}F g$, and Maxwell equation appears when $g=1$. When spinors on manifolds are introduced and the gauge transformation is extended to  $g=U(1)\times SU(2)\times U(3)$, the standard model of hadrons described by quarks and gluons was embedded in ${\bf R}^4$ space. 

In QED, only transverse components of photons interact with leptons, and Gupta-Bleuler prescription\cite{Bleuler50} to cancel non-physical degrees of freedom was proposed. In QCD, due to non-linearity of quark-gluon interactions, cancellation of non-physical degrees of freedom of gluons was performed by Faddeev-Popov ghosts. The Faddeev-Popov procedure contained the Gribov copy problem, which shows that the gauge fixing can not be done uniquely. Restricting gauge orbits to the fundamental modular region and models satisfying BRST symmetry was proposed, and by Dyson-Schwinger approach and by lattice simulation, suppression of the gluon propagator and enhancement  of the ghost propagator in infrared region was confirmed in $SU(3)$ color symmetry.

The requirement of ghost is related to the chiral symmetry or $\gamma^5$ symmetry of Dirac fermions. Constructing renormalizable QCD without ghosts was tried by several researchers, by adding an extra dimension in the Hilbert space.

In light front holographic QCD (LFHQCD), Pauli-Lubanski relativistic spin approach was adopted and subtraction of non-physical gauge freedom outside proper light-cone was performed. In the approach of non-commutative geometry, covering of manifolds with positive chirarity and manifolds with negative chirality was performed using quaternion vectors. 

If the divergent electromagnetic self energy is expressed as $\Delta (m^2)_e=m^2 \frac{3 e^2}{2\pi}log(\Lambda^2/m^2)$, where $\Lambda$ is the cut off mass, and the photon propagator is $\frac{1}{q^2}(\frac{-\Lambda^2}{-\Lambda^2-q^2})$, electro magnetic mass difference of proton and neutron mass would become\cite{Feynman72} 
\[
\Delta(M^2)_{proton}-\Delta(M^2)_{neutron}=-1.2934{\rm MeV}.
\]
Nucleons have spin 3/2 $\Delta$ resonance states, and extension of color $SU(3)$ to $SU(6)_{color, spin}$ was tried to explain hadron spectroscopy \cite{DX64} but the model was unsuccessful.  

In the Nahm-ADHM formalism in which quaternion vectors on ${\bf S}^3$ was adopted, energy of zero momentum modes were investigated to fix the scale of glueball mass\cite{vanBaal95, PvB94}.   

Jaffe \cite{Jaffe77} proposed combination of flavor $SU(3)$ (no charm contribution) and spin $SU(2)$ produces flavor singet stable dihyperon states. Incorpolation of quark and gluon degrees of freedom in deuteron form factor was done in \cite{BJL83}. Prediction of tetraquark state spectroscopy in LFHQCD is given in \cite{Brodsky19}. 

In nature, the chiral symmetry and self-adjoint operations play important roles and the quternion $\bf H$ is a useful tool, which clarifys the role of an extra dimension and conformal transformations.  Non-linearity and solitonic waves in various domains are also important\cite{FdS18}. 

\newpage
\begin{center}
{\bf Acknowledgement}
\end{center}
I thank Prof. Stanley Brodsky for presentation of AdS/QCD in 2006, Prof. Amand Faessler for instracting quark-gluon dynamics from 1981 to 1987, the late Prof. Konrad Bleuler for attracting my attention in 1989 to works of Prof. Alain Connes, when I was a research fellow of Prof. Max Huber at Bonn University, the late Prof. Hideo Nakajima for collaboration on lattice QCD simulation from 1999 to 2007. Thanks are also due to Dr. Naoki Kondo at Teikyo University, Dr. Serge Dos Santos at INSA in Blois for communications, Dr. Marc Chemtob at PTI at Saclay for the instruction from 1973 to 1974, and Libraries of Tokyo Institute of Technology for allowing consultation of references.

\vskip 0.5 true cm

\end{document}